\documentclass[onecolumn,showpacs,amsmath,amssymb,nofootinbib]
{revtex4}
\usepackage{amsmath}

\usepackage{physics}
\usepackage{subfigure}
\usepackage[T1]{fontenc}
\usepackage[utf8]{inputenc}
\usepackage{graphicx,dcolumn,bm,color,latexsym,amssymb}
\usepackage{import}
\usepackage{xifthen}
\usepackage{pdfpages}
\usepackage{transparent}
\usepackage{verbatim}

\definecolor{Blue}{rgb}{0.0,0.0,1}
\definecolor{Red}{rgb}{1,0.0,0.0}
\definecolor{Green}{rgb}{0,0.5,0.0}

\begin{document}
\title{Probabilities for informational free lunches in stochastic thermodynamics}
\author{Pedro V. Paraguass\'{u}}
 \email{paraguassu@aluno.puc-rio.br}
\affiliation{Departamento de F\'{i}sica, Pontif\'{i}cia Universidade Cat\'{o}lica\\ 22452-970, Rio de Janeiro, Brazil}
\author{Lucianno Defaveri}
\affiliation{Departamento de F\'{i}sica, Pontif\'{i}cia Universidade Cat\'{o}lica\\ 22452-970, Rio de Janeiro, Brazil}
\author{Silvio M. Duarte Queirós}\thanks{On leave of absence from: Centro Brasileiro de Pesquisas Físicas, Rio de Janeiro - RJ, Brazil}
\affiliation{National  Institute of Science and Technology for Complex Systems, Brazil}
\author{Welles A.~M. Morgado}
 \email{welles@puc-rio.br}
\affiliation{Departamento de F\'{i}sica, Pontif\'{i}cia Universidade Cat\'{o}lica\\ 22452-970, Rio de Janeiro, Brazil\\ and National  Institute of Science and Technology for Complex Systems}

\date{December 2021}

\begin{abstract}
By considering an explicit nonequilibrium model, we analyze the statistics of the irreversible work, $w_{\rm irr}$, and irreversible entropy production, $\Delta_i s$, within the stochastic energetics framework. Restating the second law of thermodynamics as a function of $w_{\rm irr}$, we introduce the explicit probability of violating the canonical form of that second law for a different set of parameters and initial conditions of the model. Moreover, we study the irreversible entropy production along the same lines, since it can be cast as a generalization of the irreversible work.
From an informational perspective, our result allows quantifying the probability of deleting information without performing work, contrarily to the Landauer's Principle, which we classify as an informational free lunch. We chose for initial conditions cases of low information content (equilibrium) and high information content (delta distributed).  
\end{abstract}

\maketitle

\section{Introduction}

The second law of thermodynamics -- namely, in the form of the Clausius inequality -- introduces the entropy as a thermodynamic state function, $S$, and establishes its inevitable increase with time for isolated systems (until a maximal is reached) and that ultimately defines which processes are reversible and which are not~\cite{callen}. With that, the possibility of setting forth a perpetual motion machine of second kind is discarded as well.
Initially related to heat -- in a differential way -- the entropy was first associated with information by Maxwell~\cite{maxwell} a link that was later explored by Szilard~\cite{szilard} with his engine and wherefrom the Landauer's principle~\cite{landauer} that is impossible to delete information without performing work can be derived.

With the advent of experimentally accessing ever smaller systems, the role of fluctuations -- negligible in canonical Thermodynamics -- has gained prominence and enhanced the probabilistic quantification of these systems \cite{paraguassu2021heat,paraguassu2021heat2,paraguassu2021heat3,chatterjee2010exact,speck2011work,ciliberto2013heat,ryabov_work_2013} and previously impossible events, viz. the so-called `free-lunches'~\cite{pellegrini2019,jarzynski2006,peliti2021stochastic}, also known as rare events, like the achievement of (very) negative values of the work, $w$, -- in the usual physical convention of work done on the system -- that dominate the averages over the Jarzinsky factor~\cite{jarzynski}, $\exp \left[ - \beta \, w \right]$, when we compress a gas at temperature, $T  \equiv \beta ^{-1}$, in a piston at any speed. Those work values are associated with negative figures of the entropy. That phenomenon is violating the canonical statement of the second law of thermodynamics, which was experimentally verified in biological~\cite{crooksrna}, purely physical~\cite{ciliberto2005} among other systems (see Ref.~\cite{ciliberto2017} for an extensive review) including non-equilibrium protocols in space like the in cascade processes that occur in fully developed turbulence~\cite{turbulence}.

In the same context, and bearing in mind the intimate relation between information and thermodynamical entropies, it is possible to look at the Landauer's principle from a probabilistic perspective and therefore by considering an informational process repeated several times for which it is possible to delete information without performing work (for very few cases). We shall coin these events Informational Free Lunches (IFL).

In the present work, we consider a thermo-mechanical overdamped system subjected to springs representing a potential in which a protocol is applied to. With that, we want to understand the conditions of the system (in terms of the values of the potential) that improve the chance of violating the second law in the fluctuating regime and whether it is possible to optimize the choice of parameters to help observe thermodynamical free lunches. These results will then clarify to what extend both the second law of Thermodynamics and the Landauer's principle are obeyed in the averaged regime.

The remaining of this manuscript is organized as follows: In Sec.~\ref{model}, we introduce our thermo-mechanical model and specify its thermodynamical properties in Sec.~\ref{secwirr} we describe the probabilistic features of the irreversible work and we link them  with the Landauer's Principle exploring the role of the initial condition in the probability space on the outcome of the irreversibility work and in Sec.~\ref{entropyproduction} the irreversible entropy production is studied along the same lines. Conclusions and final remarks are addressed to Sec.~\ref{conclusion}.

\section{Model and Thermodynamics}
\label{model}

Our starting point is an overdamped Langevin equation with a harmonic force and a driven force,
\begin{equation}
    \gamma \dot x(t) = - k x(t) + F(t) + \eta(t),\label{langevin}
\end{equation}
which describes the movement of a Brownian particle diffusing in a fluid. The thermal noise $\eta(t)$ is defined by $\langle\eta(t) \rangle=0$, and $\langle \eta(t)\eta(t')\rangle = 2\gamma \beta^{-1}\delta(t-t')$, where $\beta$ is the inverse temperature, $T$. $F(t)$ is the driving force, and here we assume that follows a linear protocol \cite{joubaud2007fluctuation},
\begin{equation}
    F(t) = F_0 \frac{t}{\tau}.
\end{equation}

The model above is simple enough but exhibits an interesting, and useful, physics that can be studied in fine detail. In special, we can analyze analytically its thermo-statistical properties such as distinct versions of the apparent violation of the second law.

The studied system has a transition probability $P[x_\tau,\tau|x_0,0]$ already calculated in the literature \cite{wio2013path,chaichian2018path} given by 
\begin{eqnarray}
  P[x_\tau,\tau|x_0,0] & = &  e^{\frac{k \tau }{\gamma }} \sqrt{\frac{k\beta}{2\pi}\left(e^{\frac{2 k \tau }{\gamma }}-1\right)^{-1}} \, \exp\left[-\frac{\beta\,\text{csch}^2\left(\frac{k \tau }{\gamma }\right)}{ 8   k^3 \tau ^2} \right. 
  \left(1 -  e^{-\frac{2 k \tau }{\gamma }}\right) \nonumber \\
   &  & 
  \left. \times   \left(\gamma  F_0-e^{\frac{k \tau }{\gamma }} (\gamma  F_0+k \tau  (k x_\tau-F_0))+k^2 \tau  x_0\right)^2\right]
%
\label{transitional}
\end{eqnarray}
which gives the probability of the particle being in the position $x_\tau$ at time $t=\tau$ given that it starts in $x_0$ at time $t=0$. That probability can be computed by path integrals methods \cite{wio2013path,suassuna2021path} or via Fokker-Planck equation~\cite{risken1996fokker}.

Due to the presence of thermal noise, the stochastic thermodynamics of the system is well defined~\cite{peliti2021stochastic,seifert2012stochastic}. Following the approach in Ref.~\cite{sekimoto2010stochastic}, we define the heat exchanged between the surrounding fluid and the particle in the time interval $t\in[0,\tau]$ as
\begin{equation}
    q[x] = \int_0^\tau (\eta(t)-\gamma \dot x) \dot x dt =  \frac{k}{2}\left(x_\tau^2-x_0^2\right) - \int_0^\tau F(t) \dot x \;dt,\label{heat}q\end{equation}
where we have used Eq.~(\ref{langevin}) to rewrite the heat. The change in the internal energy of the system during the time interval is
\begin{equation}
    \Delta U = \frac{k}{2}\left(x_\tau^2-x_0^2\right) - \left(F(\tau)x_\tau-F(0)x_0\right).
\end{equation}
Hence, we do not consider kinetic terms due to the overdamped nature of Eq.~(\ref{langevin}).
We can rewrite the heat in terms of the internal energy by integrating by parts the last term in Eq.~(\ref{heat}), which yields the first law of stochastic thermodynamics,
\begin{equation}
    q[x] =  \Delta U + \int_0^\tau \dot F(t) x(t) dt, \label{firstlaw}
\end{equation}
and allow us to obtain the definition of work \cite{jarzynski}
\begin{equation}
    w[x] \equiv - \int_0^\tau \dot F(t) x(t) dt= - \frac{F_0}{\tau}\int_0^\tau x(t) dt.
\end{equation}
The above formula is the driven work made by an external agent acting in the time interval $t\in[0,\tau]$.

Within the context of this paper, another interesting thermodynamical quantity is the irreversible work, $W_{\rm irr}$, defined as
\begin{equation}
    W_{\rm irr} \equiv \langle w[x]\rangle - \Delta F^{\rm eq},
\end{equation}
where $\Delta F^{\rm {eq}}$ is the equilibrium free energy difference. The above quantity is not a fluctuating quantity, because it only depends on the averaged quantity $\langle w[x]\rangle$ and $\Delta F^{\rm {eq}}$. That quantity obeys the irreversible work inequality version of the second law of thermodynamics when the system starts and ends in an equilibrium configuration
\begin{equation}
W_{\rm irr}\geq 0.
\label{wirrgeq0} 
\end{equation}
However, for the case where we have a nonequilibrium distribution at the end or the beginning of the process, the second law can be generalized by the Landauer's principle~\cite{esposito2011second}
\begin{equation}
W_{\rm irr}\geq \frac{\Delta I}{\beta},
\label{wirrgeqDI}
\end{equation}
where $\Delta I$ is the change in the relative entropy between the nonequilibrium distribution and the equilibrium counterpart, that is
\begin{equation}
I = \sum_x P(x)\ln\,(P(x)/\rho_0(x))= D[P(x)|| \rho_0(x)],
\end{equation}
where $D[\ldots]$ is the Kullback-Leibler distance
\cite{kullback1951information}.

\section{Fluctuating Irreversible Work}
\label{secwirr}

The second law of Thermodynamics, Eq.~(\ref{wirrgeq0}), and the generalization of the Landauer's principle, Eq.~(\ref{wirrgeqDI}), are fixed relations, since the average irreversible work, $W_{\rm irr}$ is a non-fluctuating quantity. Here, we aim at investigating the fluctuations of this irreversible work. With that goal in mind, we define the corresponding fluctuating quantity 
\begin{eqnarray}
    w_{\rm irr} \equiv w[x] -\Delta F^{\rm eq},
   \nonumber \\ = w[x] + \frac{F_0^2}{2k}
\end{eqnarray}
which has a normal probability distribution (this probability is obtained along the same lines of the probability of irreversible entropy showed in Appendix \ref{app_pi})
\begin{eqnarray}
    P(w_{\rm irr}) &=& \frac{1}{\sqrt{2\pi\sigma_w^2}} \exp\left(-\frac{(w_{\rm irr}-\mu_w)^2}{2\sigma_w^2}\right),\\
    \mu_w &=& -\frac{2 \pi  \gamma  F_0^2 \left(\gamma +\gamma  \left(-e^{-\frac{k \tau }{\gamma }}\right)-k \tau \right)}{k^3 \tau ^2},\\ \sigma_w^2 &=& \frac{2 \pi  \gamma  F_0^2 \left(\gamma +\gamma  \left(-e^{-\frac{k \tau }{\gamma }}\right)-k \tau \right) }{\beta  k^6 \tau ^4}\left(\left(1-4 \pi ^2\right) \beta  \gamma  F_0^2 \left(\gamma +\gamma  \left(-e^{-\frac{k \tau }{\gamma }}\right)-k \tau \right)-2 k^3 \tau ^2\right).
\end{eqnarray}

\begin{figure}
    \centering
    \includegraphics[width=8.6cm]{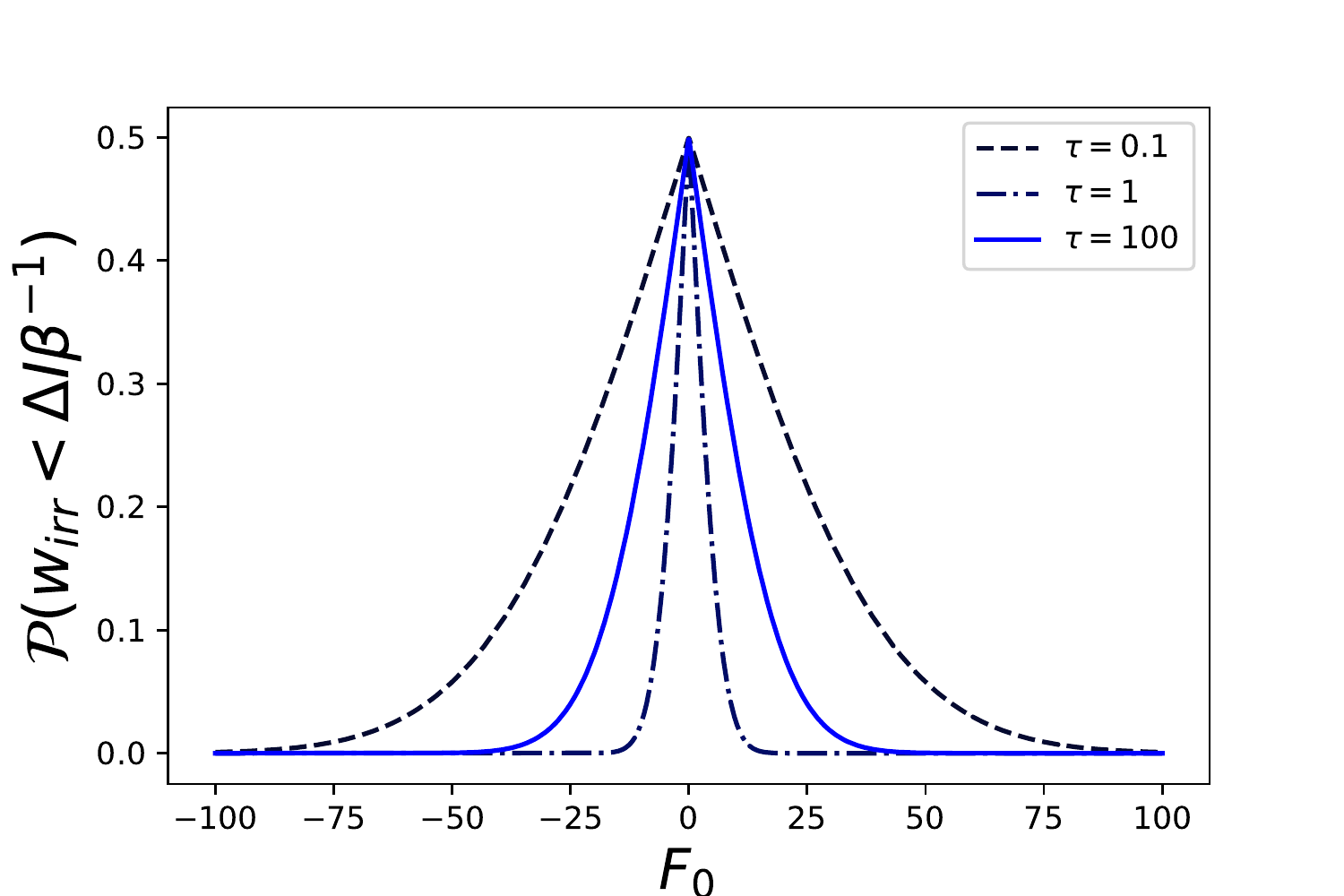}
    \caption{Probability of violating Landauer's Principle vs $F_0$ with initial equilibrium distribution. All remaining constants were set to 1.}
    \label{fig1}
\end{figure}

The relative entropy with the initial distribution being the equilibrium one, is given by
\begin{equation}
    {\Delta I}=\frac{\beta  \gamma ^2 F_0^2 e^{-\frac{2 k \tau }{\gamma }} \left(e^{\frac{k \tau }{\gamma }}-1\right)^2}{2 k^3 \tau ^2},
\end{equation}
while the probability to have $w_{\rm irr} < \Delta I \beta^{-1}$ is
\begin{equation}
    \mathcal{P}(w_{\rm irr} < \Delta I \beta^{-1}) =\int_{-\infty}^{\Delta I\beta^{-1}} P(w_{\rm irr}) dw_{\rm irr} =\frac{1}{2}\left(1+ \text{erf}\left(\frac{\beta^{-1}\Delta I - \mu_w}{\sqrt{2\sigma_w^2}}\right)\right).\label{eqinicial}
\end{equation}
This probability is plotted in Fig.~\ref{fig1}. The limiting probability 1/2 is a known result~\cite{Salazar2021}, when a detailed fluctuation theorem for the entropy production is valid. We see that the  probability to violate is larger in the vicinity of $F_0 \approx 0$. This means that a stronger protocol will make it harder to detect unusual fluctuations off the Landauer bound. Therefore, in an experimental setup, it is preferable to choose small values of this protocol. Given that the equilibrium form for the distribution corresponds to the least informational case, we may also exploit its opposite, the most informational case, i.e., the delta distributed initial condition.

\begin{figure}
    \centering
    \includegraphics[width=8.6cm]{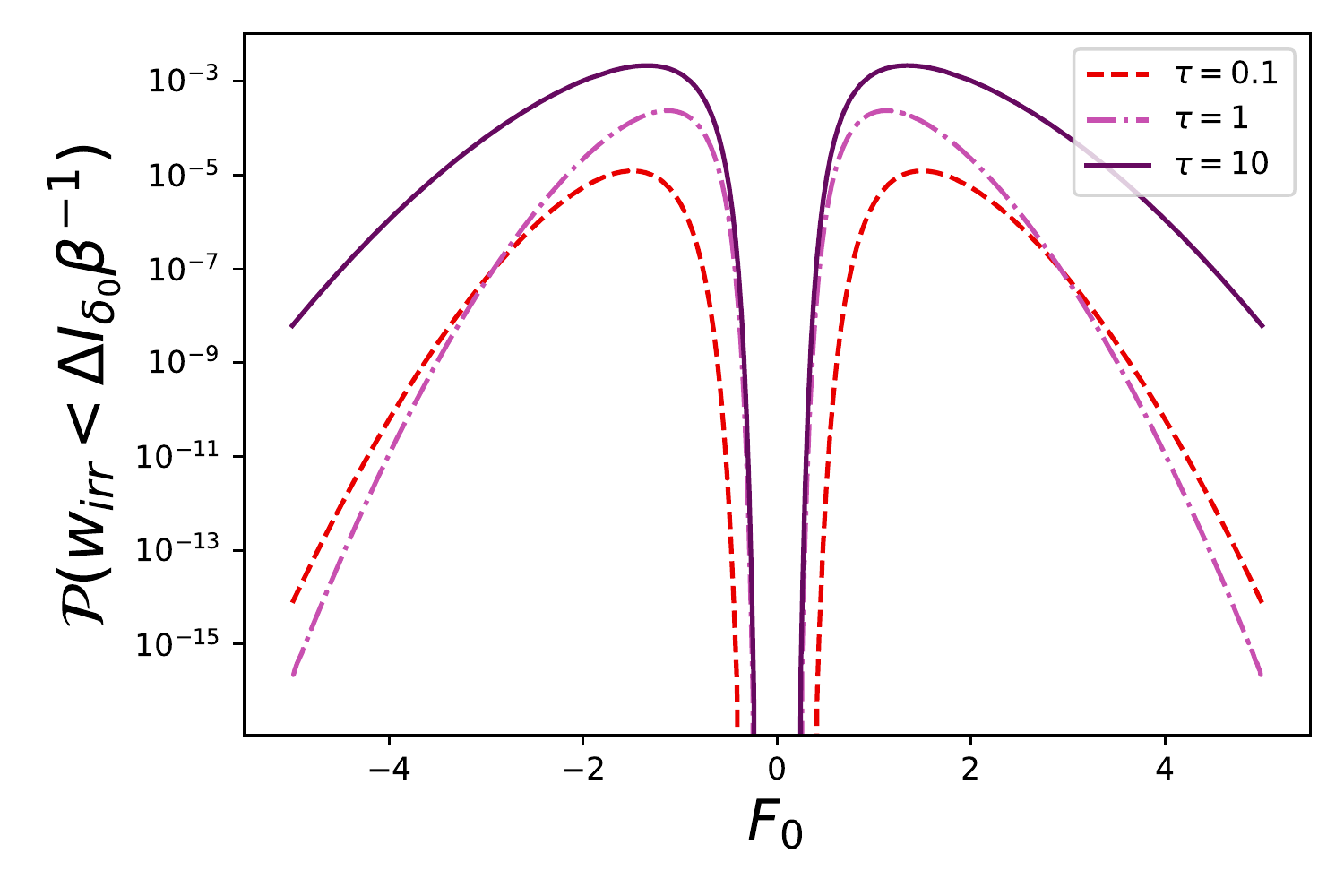}
    \caption{Probability of violating Landauer's Principle vs $F_0$ with Dirac delta distribution. Note the difference of the scales in the probabilities. All remaining constants were set to 1. }
    \label{fig2}
\end{figure}

The same calculation is possible for the case where the initial distribution is different from the equilibrium one. By choosing a Dirac delta as the initial distribution, $\rho_0(x_0)$, we focus on the case where we have the maximum knowledge about the particle position, which can be seen as a limit case of the free particle distribution
\begin{equation}
P(x_0,0) = \lim_{t\rightarrow 0 } \frac{1}{\sqrt{4\pi\gamma T t}}\exp[-\frac{x_0^2}{4\gamma Tt}] = \delta(x_0).
\end{equation}
By changing the initial distribution, we also change the relative entropy and the probability distribution of fluctuating irreversible work changes, however, the distribution is still Gaussian, changing only the mean and variance. The relative entropy in the Dirac delta case is
\begin{eqnarray}
    \Delta I_{\delta_0} &=& \frac{\beta  e^{-\frac{2 k \tau }{\gamma }} \left(F_0 e^{\frac{k \tau }{\gamma }} (2 \gamma -k \tau )-2 \gamma  F_0\right)^2}{4k^3 \tau ^2}- e^{-\frac{2 k \tau }{\gamma }} + \nonumber\\
   &+& \ln \sqrt{\frac{\beta  k \left(\coth \left(\frac{k \tau }{\gamma }\right)-1\right)}{4 \pi }}+\frac{ k \tau - \gamma}{\gamma }+1- \coth ^{-1}(3),
\end{eqnarray}
where we set $\lim_{\delta\rightarrow0}\int \delta \ln \delta =0$ by hand, because of the well known problems with continuous probabilities \cite{feller1957introduction}.
Being the work distribution Gaussian, the changes in the probability of violating the  Landauer principle come from the mean and variance. Therefore we have
\begin{eqnarray}
     \mathcal{P}(w_{\rm irr} < \Delta I_{\delta_0} \beta^{-1}) &=& \frac{1}{2}\left(1+ \text{erf}\left(\frac{\beta^{-1}\Delta I_{\delta_0} - \mu_\delta}{\sqrt{2\sigma_\delta^2}}\right)\right),\label{deltainicial}\\
     \mu_\delta &=& -\frac{F_0^2 \left(\gamma ^2-4 \gamma ^2 e^{\frac{k \tau }{\gamma }}+e^{\frac{2 k \tau }{\gamma }} (3 \gamma +k \tau ) (\gamma -k \tau )\right) e^{-\frac{2 k \tau }{\gamma }}}{2 k^3 \tau ^2},
     \\ 
     \sigma_\delta^2 &=& -\frac{\gamma  F_0^2 \left(\gamma -4 \gamma  e^{\frac{k \tau }{\gamma }}+7 \gamma  e^{\frac{2 k \tau }{\gamma }}-8 \gamma  e^{\frac{3 k \tau }{\gamma }}+2 e^{\frac{4 k \tau }{\gamma }} (2 \gamma -k \tau )\right) e^{-\frac{4 k \tau }{\gamma }}}{\beta  k^3 \tau ^2},
\end{eqnarray}
and the result is plotted in Fig.~\ref{fig2}. Now for $F_0=0$ we have no chance at all to violate the Landauer principle, since $\mathcal{P}=0$. The only chance to tune the parameters is in the two regions around zero. Interestingly, even being different plots, Figs.~\ref{fig1}~and~\ref{fig2}, in practice, inform the same thing. To have more violations in the experiment, or to have the informational free lunches, it is better to choose values of the protocol close to zero. Furthermore, in the limit $\tau\rightarrow\infty$ the probabilities have a different behaviour. For the equilibrium case, we have $\mathcal{P}\rightarrow\frac{1}{2}$, while with the Dirac delta we have $\mathcal{P}\rightarrow 0$. As we see above, there is a non-zero probability of erasing information for free, i.e., without having to realize work corresponding to $\beta^{-1}/2$ per bit erased. If care is taken to reduce the entropy produced in the process, the irreversible work can become negative~\cite{esposito2011second}.

Contrasting the results of Eqs.~\ref{eqinicial} and~\ref{deltainicial}, we observe that starting from a high entropy state yields a somewhat large probability of exacting free information (see figure~\ref{fig1}), while starting from a high information baseline renders it very difficult to extract information from the reservoir (see figure~\ref{fig2}).

\section{Fluctuating Irreversible Entropy Production}
\label{entropyproduction}
\begin{figure}
    \centering
    \includegraphics[width=8.6cm]{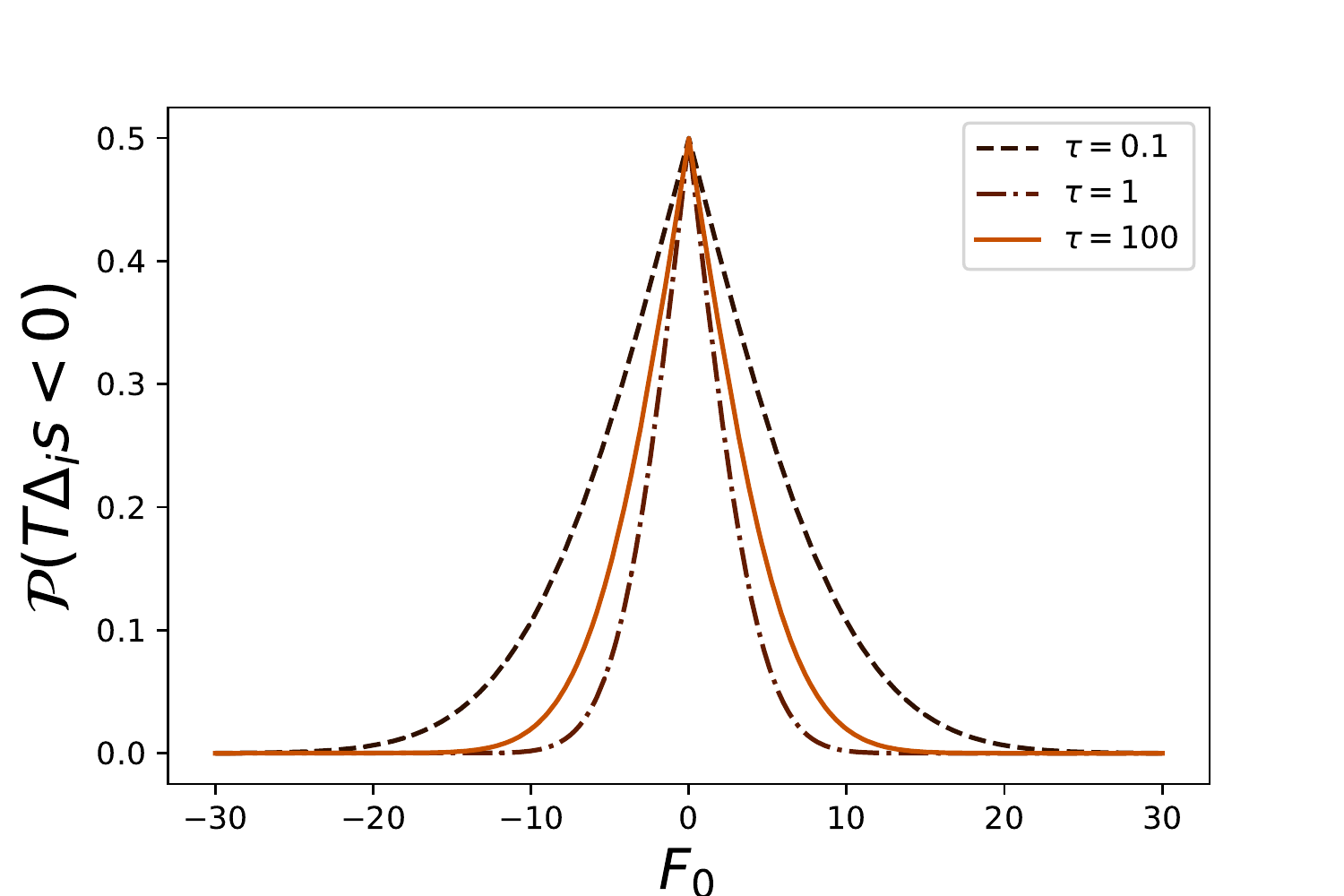}
    \caption{Probability for $T\Delta_i s <0$ vs $F_0$ with initial equilibrium distribution.  All remaining constants were set to 1}
    \label{fig3}
\end{figure}

The free energy that appears in the definition of the irreversible work is the equilibrium one. The generalization of the fluctuating irreversible work using the nonequilibrium free energy is the fluctuating irreversible entropy production, which is~\cite{esposito2011second}
\begin{equation}
    T\Delta_is[x] = w[x]-\Delta f,
\end{equation}
where $\Delta f$ is the nonequilibrium free energy, given by
\begin{equation}
    \Delta f = \Delta U - T \Delta s,
\end{equation}
with $\Delta s$ being the difference in the stochastic entropy \cite{seifert2005entropy}. The stochastic entropy can be obtained from the probability distributions for the initial and final position of the particle. Assuming an initial equilibrium distribution $\rho_0(x_0)$, and using Eq.~(\ref{transitional}) to obtain the final probability (see Appendix~\ref{app_prob}) we have
\begin{eqnarray}
   \Delta s[x] \equiv - \ln \frac{P(x_\tau,\tau)}{\rho_0(x_0)} = -\frac{1}{2} \beta  k x_0^2+\frac{\beta  {\rm e}^{-\frac{2 k \tau }{\gamma }}}{2 k^3 \tau ^2} \, \left[\gamma  F_0-{\rm e}^{\frac{k \tau }{\gamma }} (\gamma  F_0+k \tau  (k x_\tau-F_0))\right]^2.
\end{eqnarray}
Thus, the fluctuating irreversible entropy production reads 
\begin{equation}
    T\Delta_is[x] =  \frac{F_0 \, {\rm e}^{-\frac{2 k \tau }{\gamma }} }{2 k^3 \tau ^2}\left[F_0 \left(\gamma +{\rm e}^{\frac{k \tau }{\gamma }} (k \tau -\gamma )\right)^2+2 \gamma  k^2 \tau  x_\tau {\rm e}^{\frac{k \tau }{\gamma }} \left({\rm e}^{\frac{k \tau }{\gamma }}-1\right)\right]-\frac{F_0}{\tau} \int_0^\tau x(t) dt.
\end{equation}
To characterize the fluctuations of the fluctuating irreversible entropy production, we calculate its distribution which yields
\begin{equation}
    P(T\Delta_is) = \int dx_\tau \int dx_0 \rho_0(x_0) \int_{x_0,x_\tau} Dx e^{S[x]} \delta (T\Delta_is - T\Delta_is[x]).
\end{equation}
where $S[x]$ is the stochastic action, defined in Appendix~\ref{app_pi} together with the calculation of the path integral. After carrying out the integration, the result is a normal distribution for the fluctuating irreversible entropy production
\begin{eqnarray}
    P(T\Delta_is) &=& \frac{1}{\sqrt{2\pi\sigma_w^2}}\exp\left(\frac{-(T\Delta_is-\mu_w)^2}{2\sigma_w^2}\right),
\label{workdist}    \\ \mu_s &=& \frac{\gamma  F_0^2 e^{-\frac{3 k \tau }{\gamma }} \left(-\gamma +4 \gamma  e^{\frac{k \tau }{\gamma }}+e^{\frac{2 k \tau }{\gamma }} (2 k \tau -3 \gamma )\right)^{3/2}}{2 k^3 \tau ^2 \sqrt{\gamma  \left(e^{-\frac{2 k \tau }{\gamma }} \left(4 e^{\frac{k \tau }{\gamma }}-1\right)-3\right)+2 k \tau }},\;\;\; \sigma_s^2 = 2\mu_w\beta^{-1}
\end{eqnarray}
The above distribution satisfies the detailed fluctuation theorem  
\begin{equation}
    P(\Delta_is)/P(-\Delta_is) = e^{\Delta_is},
\end{equation}
which is in accordance with the second law of Thermodynamics, since $\langle T\Delta_is \rangle \equiv T\Delta_iS\geq 0$ follows from the Jensen inequality. Now, instead of Landauer's principle, we want to see how we can improve the choice of the parameters $\tau$ and $F_0$ to maximize the probability of $T\Delta_i s<0$. Therefore, the probability that the fluctuating quantity $T\Delta_i s$ has values less than zero is
\begin{equation}
    \mathcal{P}(T\Delta_is<0) = \frac{1}{2} \text{erfc}\left(\frac{|F_0| e^{-\frac{k \tau }{\gamma }} \sqrt{\beta  \gamma  \left(-\gamma +4 \gamma  e^{\frac{k \tau }{\gamma }}+e^{\frac{2 k \tau }{\gamma }} (2 k \tau -3 \gamma )\right)}}{2 \sqrt{2} k^{3/2} \tau }\right),
\end{equation}
where $ \rm erfc$ is the complementary error function. The result is plotted in Fig.~\ref{fig3}.

\begin{figure}
    \centering
    \includegraphics[width=8.6cm]{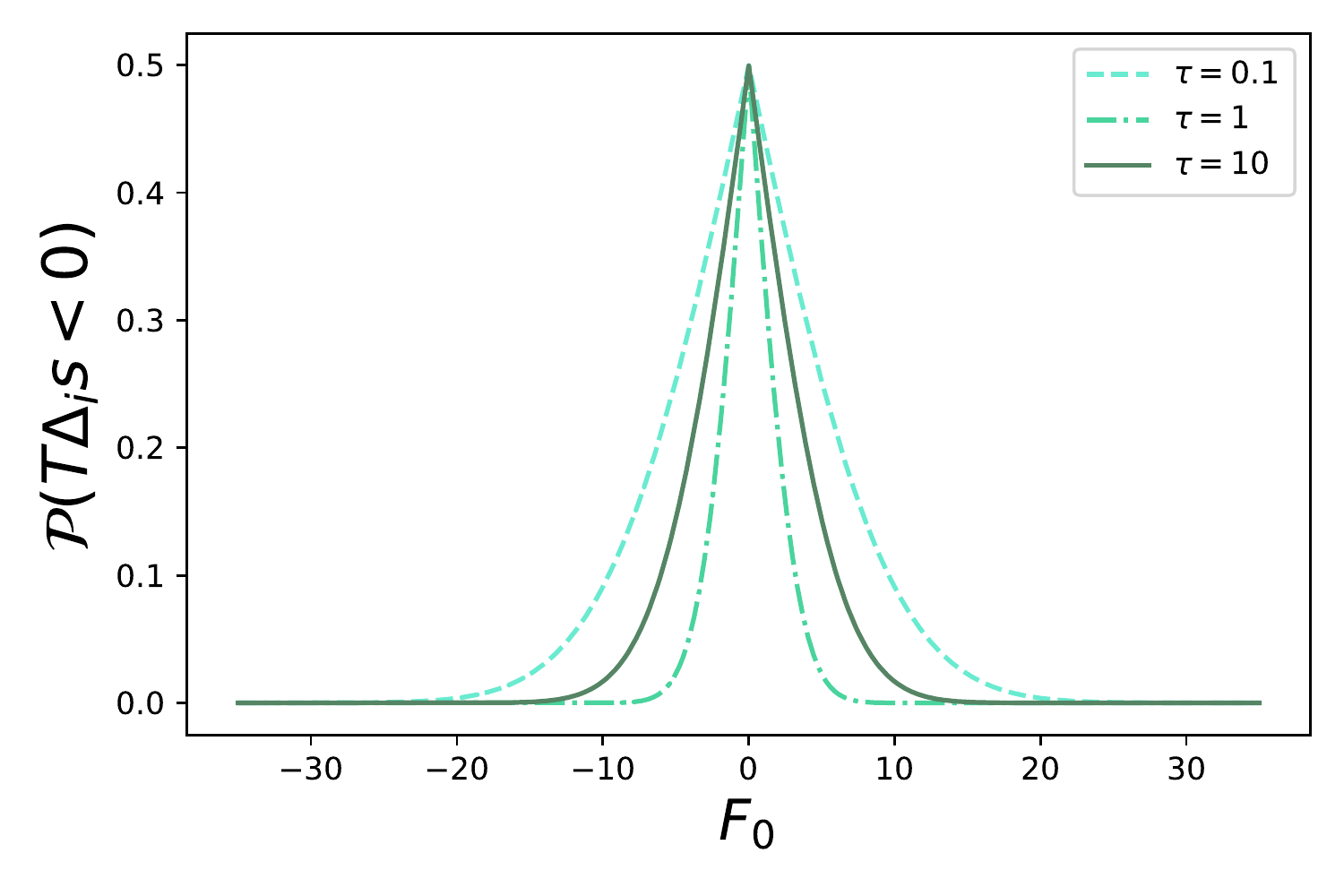}
    \caption{Probability for $T\Delta_i s <0$ vs $F_0$ with initial Dirac delta distribution.}
    \label{fig4}
\end{figure}

Moreover, we can consider the case where the Dirac Delta is the initial distribution as well. For this, we have
\begin{eqnarray}
   P_{\delta_0}(w_{\rm irr}) &=& \frac{1}{\sqrt{2\pi \sigma_\delta^2}}\exp\left[-\frac{(w_{\rm irr}-\mu_\delta)^2}{2\sigma_\delta^2}\right],
   \\ \mu_\delta &=& -\frac{\gamma  F_0^2 }{2 k^3 \tau ^2}\left[\gamma  \left(e^{-\frac{2 k \tau }{\gamma }}-4 e^{-\frac{k \tau }{\gamma }}+3\right)-2 k \tau \right],\\ \sigma_\delta^2 &=& \frac{\gamma  F_0^2 e^{-\frac{4 k \tau }{\gamma }} }{\beta  k^3 \tau ^2}\left(\gamma -4 \gamma  e^{\frac{k \tau }{\gamma }}+7 \gamma  e^{\frac{2 k \tau }{\gamma }}-8 \gamma  e^{\frac{3 k \tau }{\gamma }}+2 e^{\frac{4 k \tau }{\gamma }} (2 \gamma -k \tau )\right)
\end{eqnarray}
where we set the subscript $\delta_0$ emphasizes we are coping with a Dirac delta as the initial distribution. As expected, that distribution does not obey the detailed fluctuation theorem because we no longer have an equilibrium distribution. Nevertheless, for $\tau\rightarrow\infty$ we have the asymptotic fluctuating theorem, 
\begin{eqnarray}
    \ln \frac{P_{\delta_0}(w_{\rm irr})}{P_{\delta_0}(-w_{\rm irr})} &=& \frac{2\mu_\delta}{\sigma_\delta^2} \, w_{\rm irr},\\ \lim_{\tau\rightarrow\infty}\frac{2\mu_\delta}{\sigma_\delta^2} &=& \lim_{\tau\rightarrow\infty}\frac{\beta  e^{\frac{2 k \tau }{\gamma }} \left(\gamma -4 \gamma  e^{\frac{k \tau }{\gamma }}+e^{\frac{2 k \tau }{\gamma }} (3 \gamma -2 k \tau )\right)}{\gamma -4 \gamma  e^{\frac{k \tau }{\gamma }}+7 \gamma  e^{\frac{2 k \tau }{\gamma }}-8 \gamma  e^{\frac{3 k \tau }{\gamma }}+2 e^{\frac{4 k \tau }{\gamma }} (2 \gamma -k \tau )} = \beta.
\end{eqnarray}
The probability for $T\Delta_i s<0$ in the Dirac delta case is plotted in figure \ref{fig4}.

For the fluctuating irreversible entropy production, the behavior of the probability does not change considerably by changing the initial condition. As we can see from figures \ref{fig3} and \ref{fig4}, it is better to choose a small time together with the choice of a small value of $F_0$ to improve the probabilities of $T\Delta_i s <0$. Moreover, for $\tau\rightarrow\infty$ both probabilities become $\mathcal{P}(T\Delta_i s<0)=\frac{1}{2}$ which is the maximum value allowed and corresponds to the equilibration of the system, where the protocol is absent. 

\section{Conclusion}
\label{conclusion}

Irreversible work and entropy production obey different forms of the second law of Thermodynamics. In the present paper, we have aimed at studying how fluctuating versions of those quantities can seemingly violate the canonical form of second law, which we have coined as informational free lunches (IFL) where, specifically, it was possible to erase information without performing work, an implication that goes against the canonical statement of Landauer's principle. Interestingly, it is precisely because of these very few single trajectory cases for which one can both delete information and even receive work that Landauer's principle can be macroscopically asserted. In our approach, where we have considered an overdamped classical system, information boils down to the tracking of the particle, which can be made by means equilavent to those originally introduced for retrieving the trajectory of particles in a turbulent fluid~\cite{bodenschatz01,bodenschatz02,bodenschatz09} and wherefrom the empirical distribution functions for the computation of the stochastic produced entropy, $\Delta _{(i)} s$, are obtained.

We have learned that, for the system in question, namely the protocol used, the best way to obtain the IFLs is to consider small protocol intensity values applied during small time intervals. This behavior was observed in all cases studied, using different initial conditions; both for irreversible entropy and irreversible work. This is consistent with the fact that the production of irreversible entropy can be seen as a generalization of irreversible work; when $\tau $ is very large in comparison with the other time scale of the system $\gamma / k$ for which the protocol would be closer to a quasi-static transformation where irreversible processes are negligible.

We observe that starting from a high entropy state yields a somewhat large probability of exacting free information (see figure~\ref{fig1}), while starting from a high information baseline renders it very difficult to extract information from the reservoir (see figure~\ref{fig2}).

Understanding how to choose the parameters that improve in obtaining the IFLs is important experimentally, as these events are less likely to occur while they are needed to obtain good system statistics. Along these lines, some questions are left for future work: Can we find an optimized protocol in the harmonic case to improve the emergence of IFLs? Are the results obtained in this work still valid if we have non-harmonic potentials?

\appendix

\section{Probabilities} \label{app_prob}

From the conditional position probability we can obtain the position distribution. We just need to integrate out the conditional. Since we are assuming an equilibrium initial distribution we have
\begin{eqnarray}
    P(x_\tau,\tau)&=& \int P[x_\tau,\tau|x_0,0] \rho_0(x_0) dx_0   
    \nonumber\\ &=& \sqrt{\frac{\beta k}{2\pi}}\exp \left(-\frac{\beta  e^{-\frac{2 k \tau }{\gamma }} \left(\gamma  F_0-e^{\frac{k \tau }{\gamma }} (\gamma  F_0+k \tau  (k x_\tau-F_0))\right)^2}{2 k^3 \tau ^2}\right).
\end{eqnarray}
The above expression is used to calculate the stochastic entropy. For the case where the initial distribution is the Dirac delta, we only need to use the properties of the delta, giving
\begin{equation}
    P_{\delta_0}(x_\tau,\tau) = \int P[x_\tau,\tau|x_0,0] \delta(x_0) dx_0 =P[x_\tau,\tau|0,0]. 
\end{equation}

\section{Calculation of irreversible entropy production distribution} \label{app_pi}

To calculate the distribution of the irreversible entropy production, we need to solve the integration below
\begin{equation}
    P(T\Delta_i s) = \int dx_\tau \int dx_0 \rho_0(x_0) \int_{x_0,x_\tau} Dx e^{S[x]} \delta (T\Delta_is - T\Delta_is[x]).
\end{equation}
The first step is to rewrite the Dirac delta as an integral, giving
\begin{equation}
    P(T\Delta_i s) = \int \frac{d\lambda}{2\pi} e^{i\lambda T\Delta_i s} Z_s(\lambda),
\end{equation}
where $Z_s(\lambda)$ is the characteristic function of the irreversible entropy production given by
\begin{equation}
    Z_s(\lambda) = \int dx_\tau \int dx_0 \rho_0(x_0) \int_{x_0,x_\tau} Dx e^{S[x]-i\lambda T\Delta_i s[x]}.
\end{equation}
The dependence on the trajectory of the irreversible entropy production comes only from the work $T\Delta_i s[x] = W[x]-\Delta f$. This means that only $W[x]$ affects the path integral. Therefore, we need to solve the path integral, which is a Gaussian path integral, and the solution is
\begin{equation}
   I_W = \int_{x_0,x_\tau} Dx e^{S[x]-i\lambda w[x]}.
\end{equation}
The expression is long, but then the remaining is a Gaussian expression. After solving the path integral, it remains to solve the integrals in $x_0 $ and $x_\tau$, which are Gaussian integrals. After integrating all Gaussian integrals, we are left with the characteristic function
\begin{equation}
      Z_{w}(\lambda)= \exp \left(\frac{\gamma  F_0^2 \lambda  (\lambda +i \beta ) \, e^{-\frac{2 k \tau }{\gamma }} \left(\gamma -4 \gamma  e^{\frac{k \tau }{\gamma }}+e^{\frac{2 k \tau }{\gamma }} (3 \gamma -2 k \tau )\right)}{2 \beta  k^3 \tau ^2}\right),
\end{equation}
and we now only need to integrate over $\lambda$, which is also a Gaussian integral. The result is the probability distribution of the irreversible entropy production written in eq.~\ref{workdist}.

\end{document}